\documentclass{PoS}

\usepackage{amssymb}

\title{Using Forward-Backward Drell-Yan Asymmetry in PDF Determinations}

\ShortTitle{DY AFB for PDF determination}

\author{\speaker{Juri Fiaschi}\footnote{On behalf of the {\tt{xFitter}} developer's team.}\\
        Institut f\"ur Theoretische Physik, Westf\"alische Wilhelms-Universit\"at
 M\"unster, Wilhelm-Klemm-Stra\ss{}e 9, D-48149 M\"unster, Germany\\
        E-mail: \email{fiaschi@uni-muenster.de}}

\author{Elena Accomando\\
       School of Physics \& Astronomy, University of Southampton, Southampton SO17 1BJ, UK\\
       E-mail: \email{e.accomando@soton.ac.uk}}
       
\author{Francesco Hautmann\\
       Theoretical Physics Department, University of Oxford, Oxford OX1 3NP, UK\\
       CERN, Theoretical Physics Department, CH 1211, Geneva, Switzerland\\
       E-mail: \email{hautmann@thphys.ox.ac.uk}}

\author{Stefano Moretti\\
       School of Physics \& Astronomy, University of Southampton, Southampton SO17 1BJ, UK\\
       E-mail: \email{s.moretti@soton.ac.uk}}

\abstract{We study the impact of the inclusion of Neutral Current (NC) DY data from LHC mapped in the Forward-Backward Asymmetry ($A_{\rm FB}$) observable on PDF uncertainties, using the open source platform \texttt{xFitter}.
We find that $A_{\rm FB}$ enables new PDF sensitivity at current and future luminosity stages of LHC.}

\FullConference{XXVII International Workshop on Deep-Inelastic Scattering and Related Subjects - DIS2019\\
		8-12 April, 2019\\
		Torino, Italy}

\begin{document}
\vspace*{-0.25truecm}
\begin{flushright}
 \hspace{3cm} MS-TP-19-16\\
 \hspace{3cm} CERN-TH-2019-100\\
\end{flushright}
\vspace*{-0.25truecm}

\section{Introduction}

The current and future runs at the LHC will deliver a conspicuous amount of data, which will be used to extend searches for Beyond Standard Model (BSM) physics in the high energy region and will allow precise measurements to test the Standard Model (SM) in the low energy regime.
The Parton Distribution Functions (PDFs) are a necessary fundamental tool that is required in order to compare the data with theoretical predictions.
The precision in the determination of the PDFs is a crucial factor in the evaluation of theoretical systematics.
Improving the determination of the PDFs is therefore an important task, which will be carried out by adding new LHC data and by including additional observable that would serve the purpose.

In this work, we study the impact from the inclusion of Drell-Yan (DY) di-lepton data in the form of Forward-Backward Asymmetry ($A_{\rm FB}$) in the determination of PDFs.
It was already shown that this observable can provide constraints on the PDFs~\cite{Accomando:2018nig,Accomando:2017scx} and here we quantify this effect by using the open source platform \texttt{xFitter}~\cite{Alekhin:2014irh} for the profiling of various sets of PDFs~\cite{Abdolmaleki:2019qmq}.
We employ $A_{\rm FB}$ pseudodata with a statistical precision corresponding to different integrated luminosities ranging from 30 fb$^{-1}$ to 3000 fb$^{-1}$.

\section{Details of the analysis}

We started by implementing a suitable \texttt{C++} code for the calculation of the $A_{\rm FB}$ at Leading Order (LO) and integrating it into the \texttt{xFitter} environment.
Drell-Yan production of di-lepton pair is mediated by the exchange of a photon or a $Z$ boson, and the $A_{\rm FB}$ is defined as

\begin{equation}
  A_{\rm{FB}}^* = \frac { d \sigma / d M(\ell^+\ell^-)[\cos\theta^*>0] - d \sigma / d M(\ell^+\ell^-)[\cos\theta^*<0] }      { d \sigma / d M(\ell^+\ell^-)[\cos\theta^*>0] + d \sigma / d M(\ell^+\ell^-)[\cos\theta^*<0] }.
  \label{eq:afb}
\end{equation}

\noindent
At LO, the angle $\theta^*$ is defined according to the direction of the incoming quark, which is assumed to be given by the direction of the boost of the final state di-lepton pair~\cite{Dittmar:1996my,Rizzo:2009pu,Accomando:2015cfa,Accomando:2016tah,Accomando:2016ehi}.

Subsequently we included NLO corrections to the observable, employing suitable NLO grids obtained with {\tt{MadGraph5{\_}aMC@NLO}}~\cite{Alwall:2014hca} interfaced to {\tt{APPLgrid}}~\cite{Carli:2010rw} through {\tt{aMCfast}}~\cite{Bertone:2014zva}.
At NLO the angle $\theta^*$ is defined is defined in the Collins-Soper (CS) frame~\cite{Collins:1977iv} as

\begin{equation}
\cos\theta^{*} = \frac{p_{{Z},\ell\ell}}{M_{\ell\ell}|p_{{Z},\ell\ell}|} \frac{p_{1}^{+}p_{2}^{-}-p_{1}^{-}p_{2}^{+}}{\sqrt{M^{2}_{\ell\ell}+p^{2}_{{T},\ell\ell}}},
\end{equation}

\noindent
with $p_{i}^{\pm}=E_{i}\pm p_{{Z},i}$, the index $i = 1,2$ labeling the positive and negative charged lepton respectively. $E$ and $p_{{Z}}$ are the energy and the $z$-components of the leptonic four-momentum, respectively; $p_{{Z},\ell\ell}$ is the di-lepton $z$-component of the momentum and $p_{{T},\ell\ell}$ is the di-lepton transverse momentum. 

The cross sections appearing in Eq.~\ref{eq:afb} are calculated in the fiducial region, i.e. applying the standard kinematical cuts to simulate the detector acceptance ($|\eta_\ell| < 2.5$ and $p^{\ell}_{{T}} >$ 20 GeV).
A set of datafiles has been generated for each PDF set under analysis.
They contain the predictions for the observable in the invariant mass range $45~{\rm GeV} < M_{\ell\ell} < 200~{\rm GeV}$ with a bin width of 2.5 GeV, and the projected statistical error, obtained following

\begin{equation}
  \Delta A_{\rm{FB}}^* = \sqrt{\frac{1-{A_{\rm{FB}}^*}^2}{N}}, 
  \label{eq:afb_error}
\end{equation}

\noindent
where the number of events $N$ includes an acceptance times efficiency factor of the order of $\sim$20\%~\cite{Khachatryan:2014fba} and NNLO QCD corrections included by adopting a mass dependent $k$-factor~\cite{Hamberg:1990np, Harlander:2002wh}.
The datafiles have been generated fixing the collider centre-of-mass energy at 13 TeV and for three values of integrated luminosity (30 fb$^{-1}$, 300 fb$^{-1}$ and 3000 fb$^{-1}$).
Furthermore we also considered the effects on the PDF profiling from the application of rapidity cuts on the di-lepton system.
For this purpose, datafiles have been generated with $|y_{\ell\ell}| > 0$ (no rapidity cut), $|y_{\ell\ell}| > 1.5$ and $|y_{\ell\ell}| > 4.0$.
The profiling technique~\cite{Paukkunen:2014zia} is based on minimizing $\chi^2$ between data and theoretical predictions.
It assumes that the new data are compatible with the theoretical predictions using the existing PDF set and, under this assumption, the central values of the data points are set to the central values of the theoretical predictions.
No theoretical uncertainties except the PDF uncertainties are considered when calculating the $\chi^2$.

\section{Profiled PDFs with $A_{\rm FB}$}

\begin{figure}[h]
\begin{center}
\includegraphics[width=0.25\textwidth]{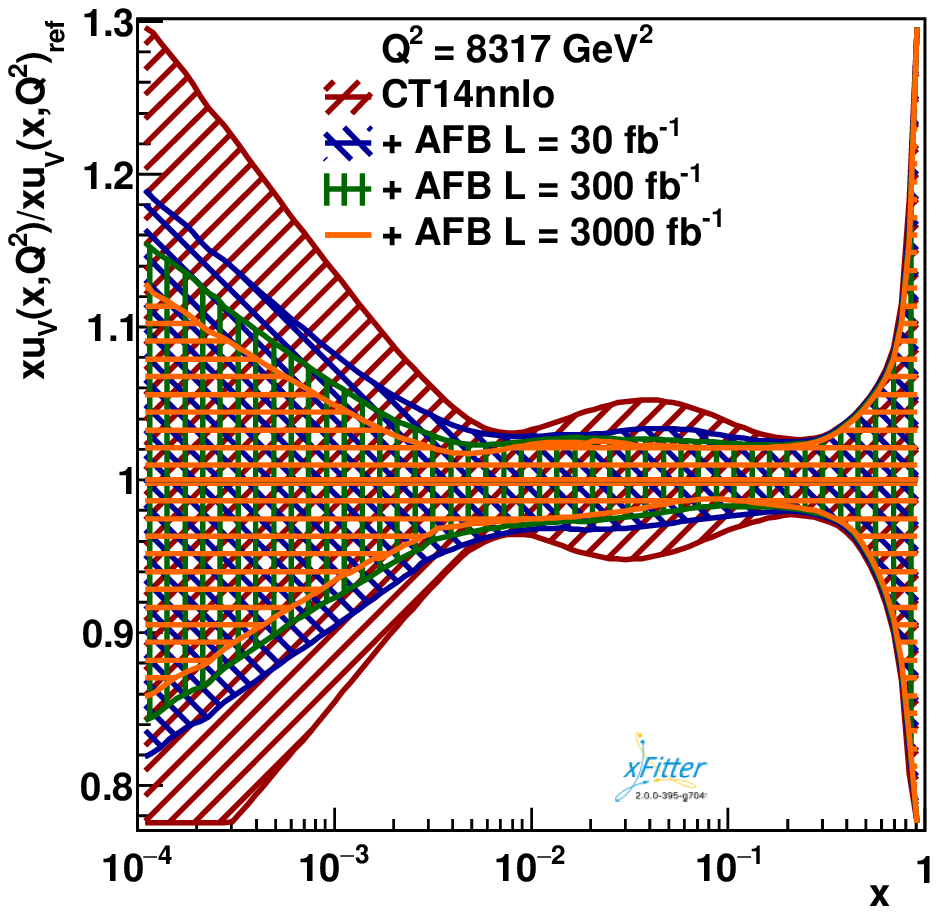}\includegraphics[width=0.25\textwidth]{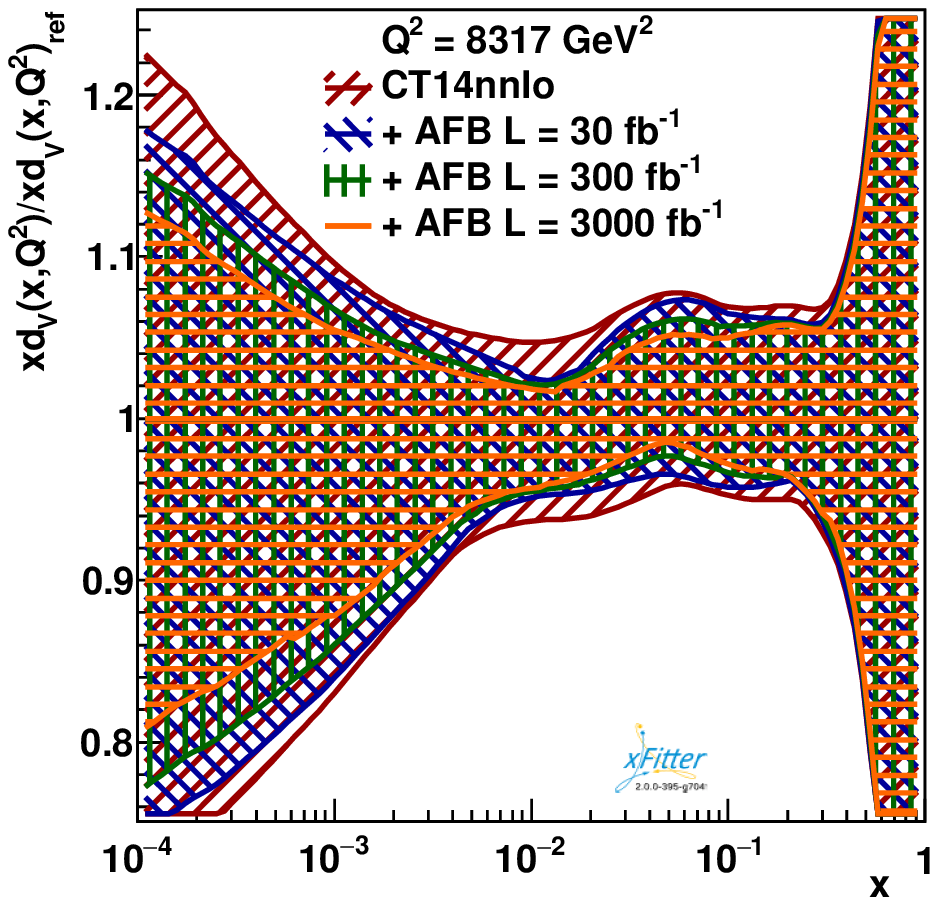}\includegraphics[width=0.25\textwidth]{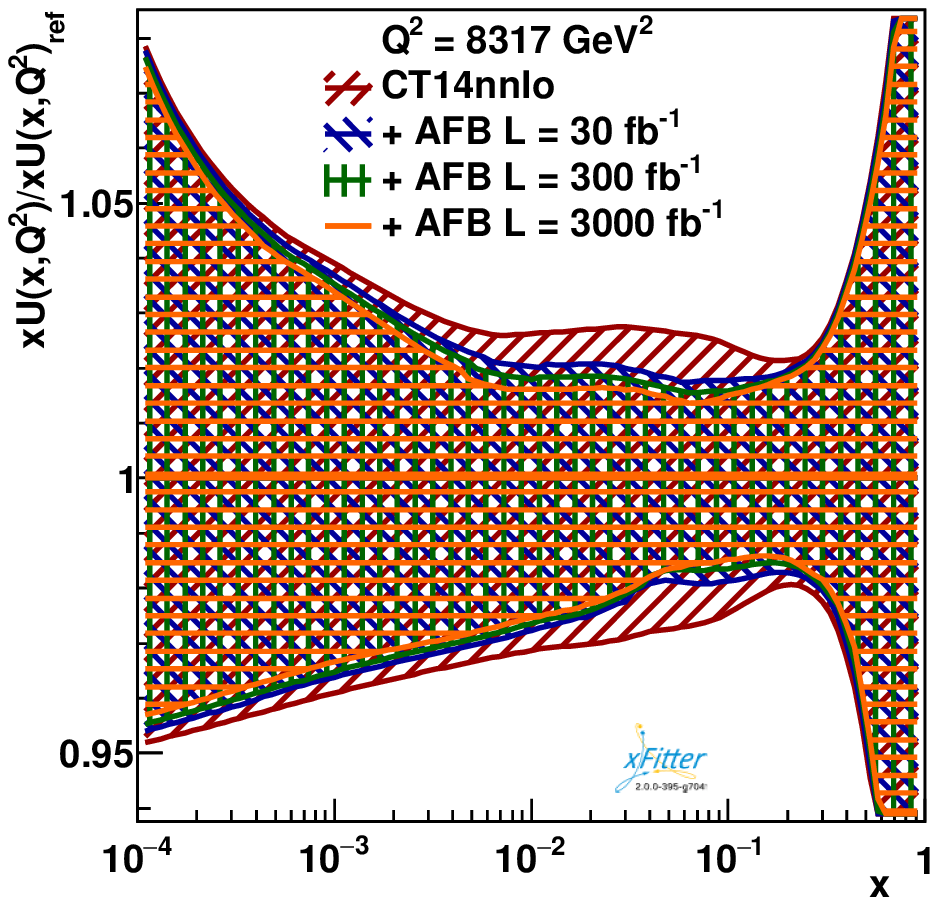}\includegraphics[width=0.25\textwidth]{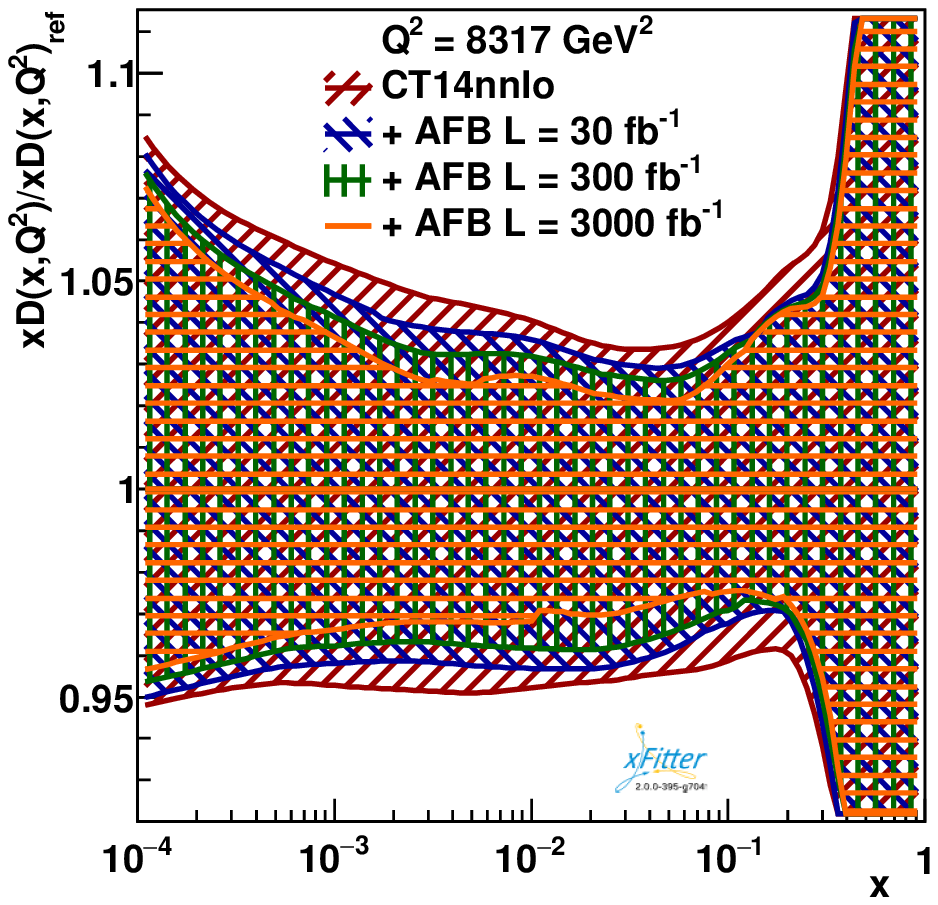}
\caption{Original and profiled distributions for the normalised ratios of (left to right) $u$-valence, $d$-valence, $u$-sea and $d$-sea quarks of the CT14nnlo PDF set, obtained with $A_{\rm{FB}}^*$ pseudodata at different luminosities.}
\label{fig:prof_CT14nnlo_lum}
\end{center}
\end{figure}

In this section we present the results for the profiling of various PDF sets when including $A_{\rm FB}$ measurements.
In Fig.~\ref{fig:prof_CT14nnlo_lum} we show the results of the profiling on the CT14nnlo~\cite{Dulat:2015mca} PDF set.
The error bands obtained with this PDF set have been rescaled to 68\% CL for a consistent comparison with the results obtained for other PDF sets.
We display the error bands for the normalised distribution of $u$-valence, $d$-valence, $u$-sea and $d$-sea quarks.
The profiled curves are obtained including $A_{\rm FB}$ data with a statistical error corresponding to integrated luminosities of 30 fb$^{-1}$ (blue), 300 fb$^{-1}$ (green) and 3000 fb$^{-1}$ (orange).
The $u$-valence and the $d$-valence quark distributions show the largest sensitivity to the new data, especially in the region of intermediate and low momentum fraction $x$.
Sea quark distributions also exhibit some improvement, particularly in the region of intermediate $x$ for high luminosity.

\begin{figure}[h]
\begin{center}
\includegraphics[width=0.25\textwidth]{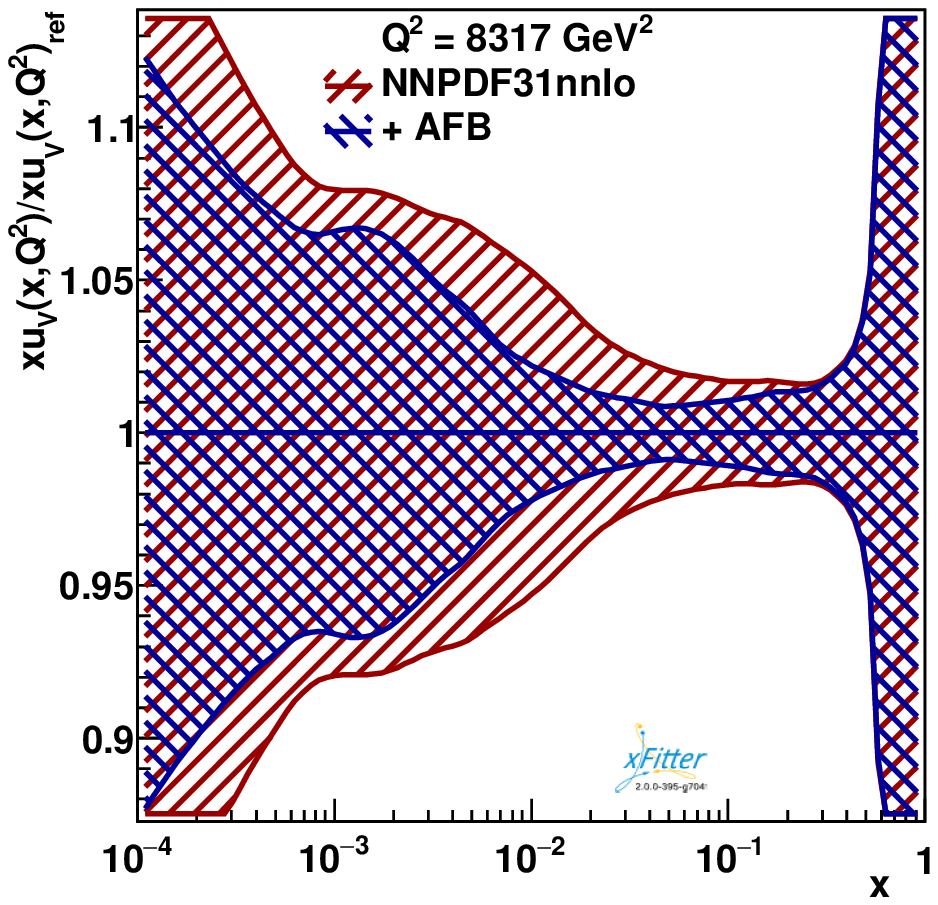}\includegraphics[width=0.25\textwidth]{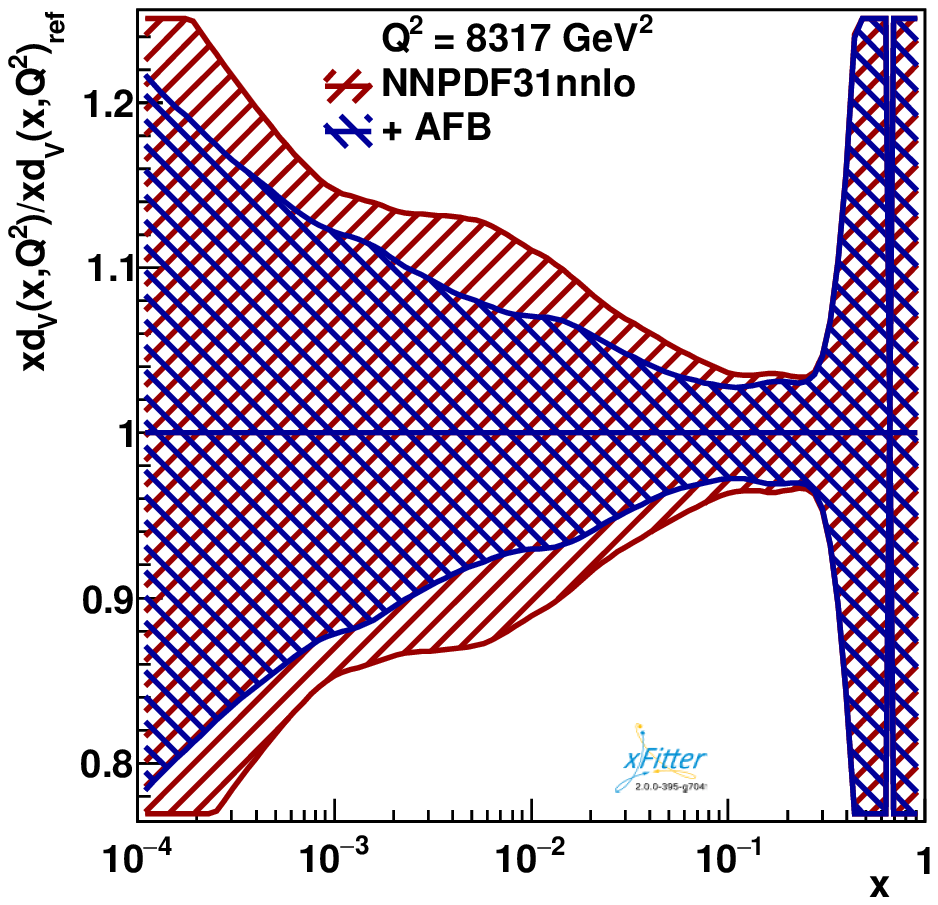}\includegraphics[width=0.25\textwidth]{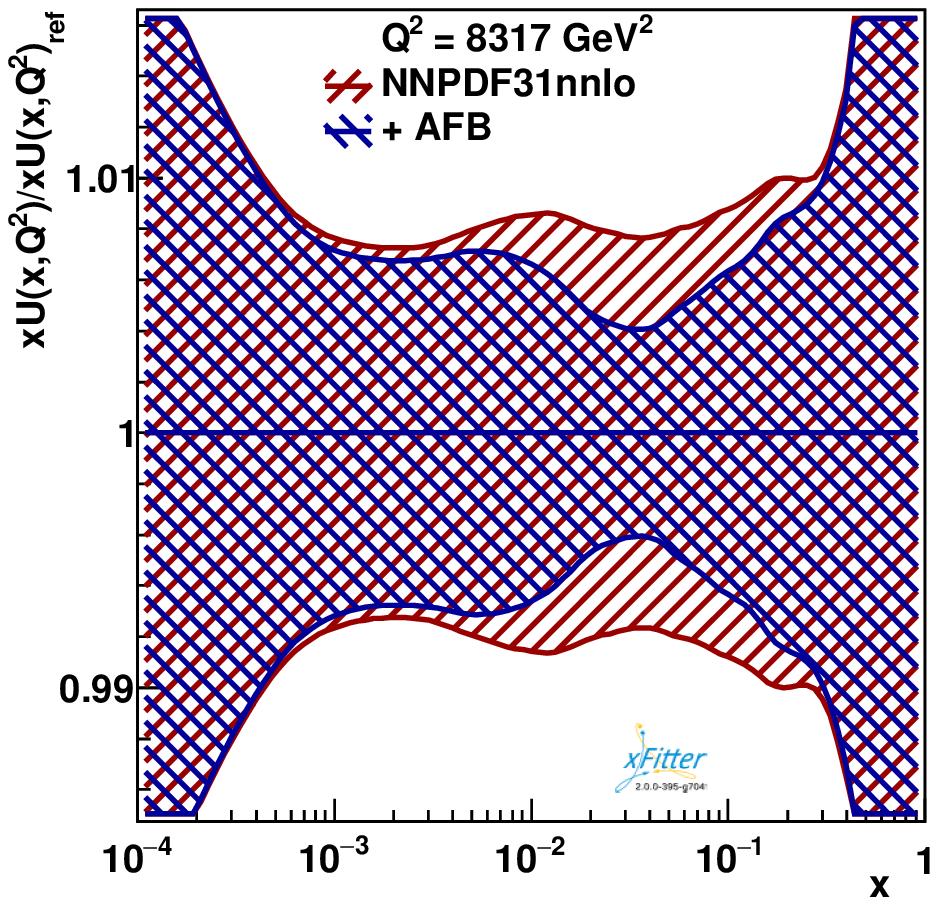}\includegraphics[width=0.25\textwidth]{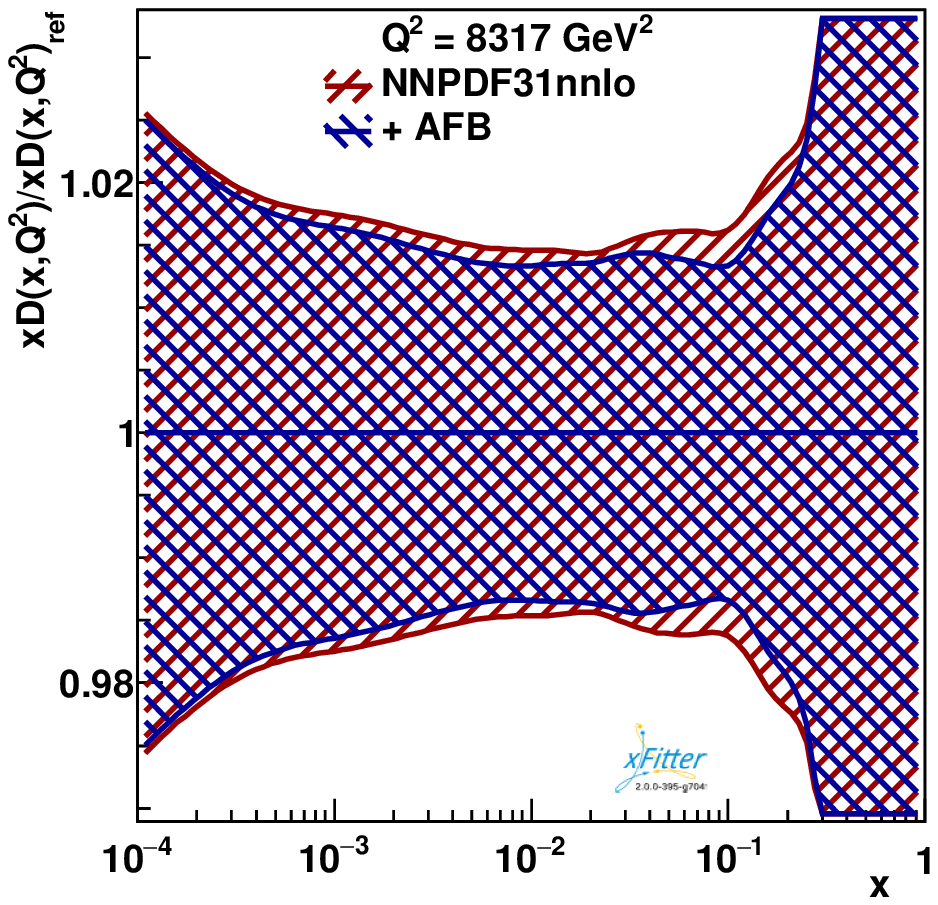}\\
\includegraphics[width=0.25\textwidth]{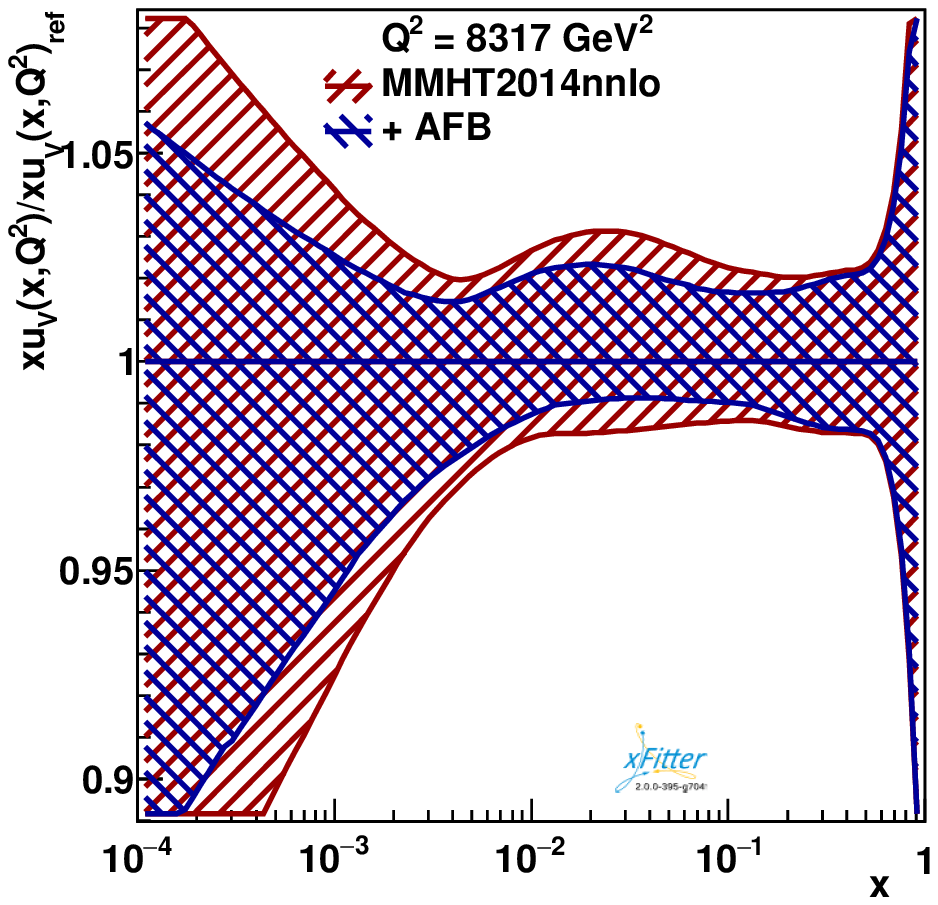}\includegraphics[width=0.25\textwidth]{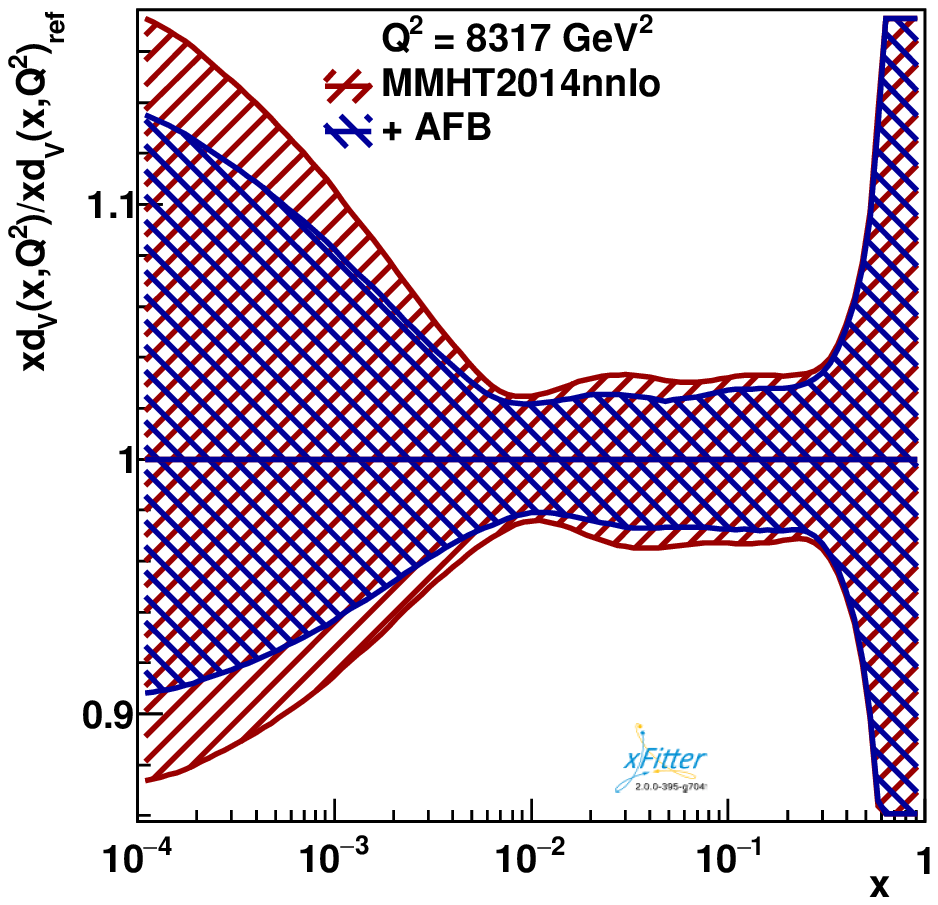}\includegraphics[width=0.25\textwidth]{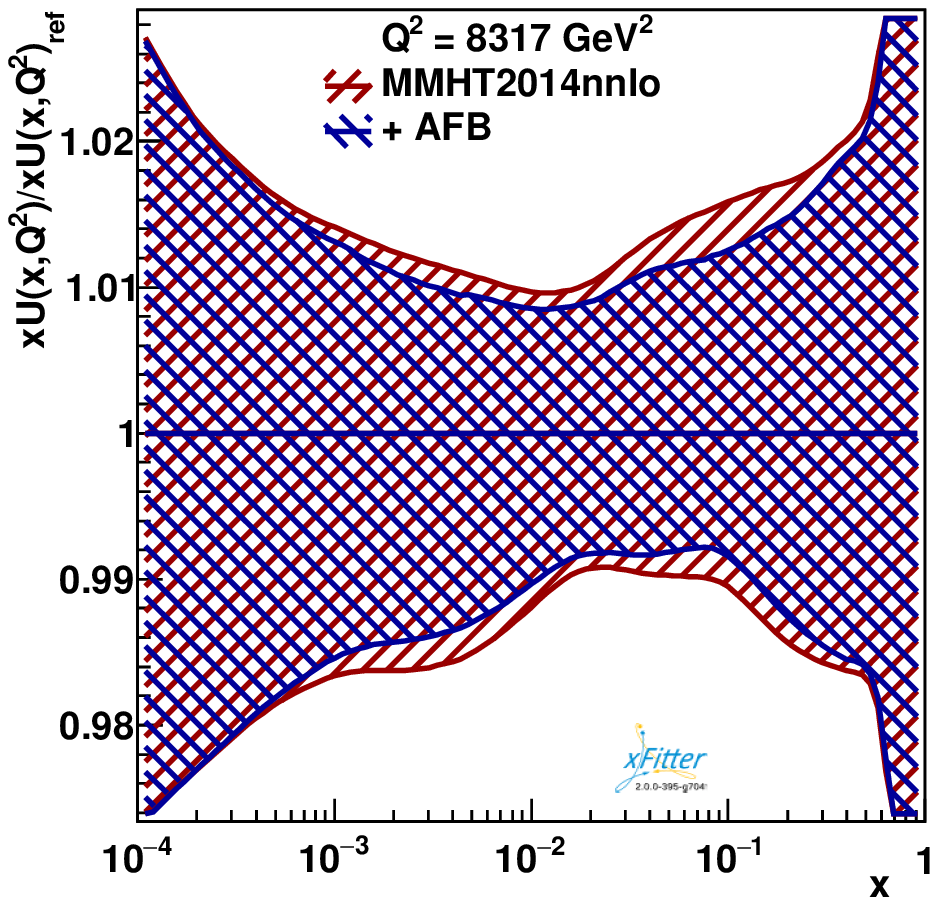}\includegraphics[width=0.25\textwidth]{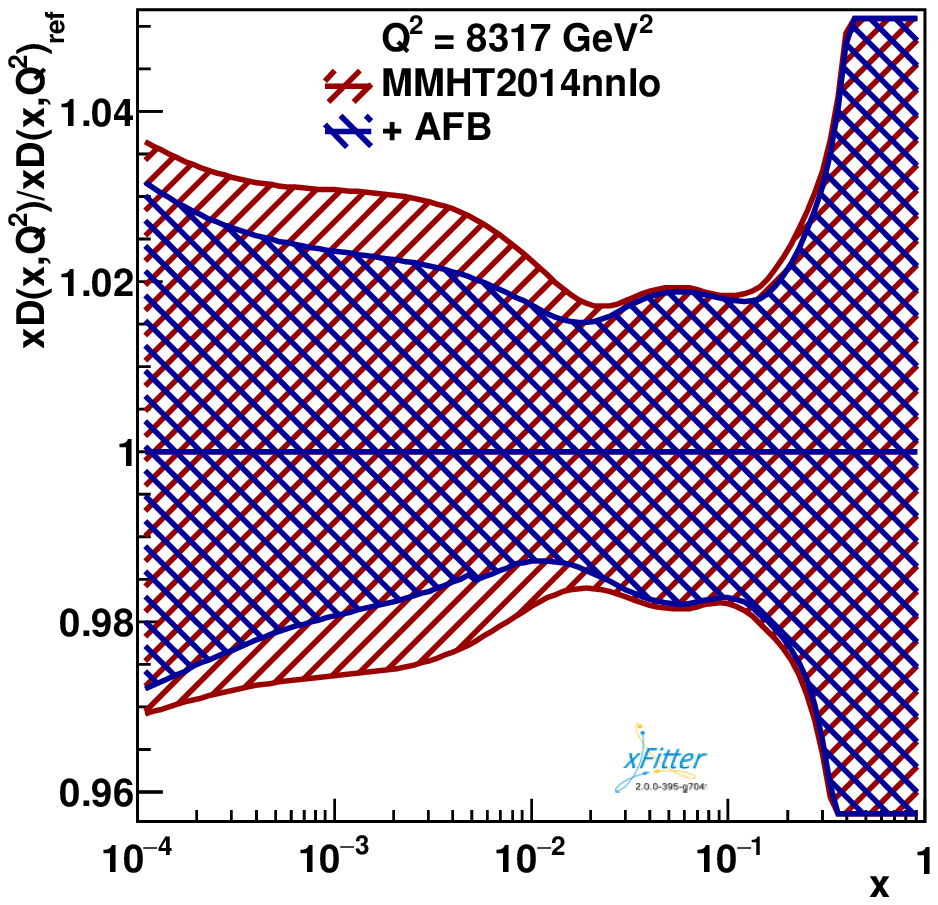}\\
\includegraphics[width=0.25\textwidth]{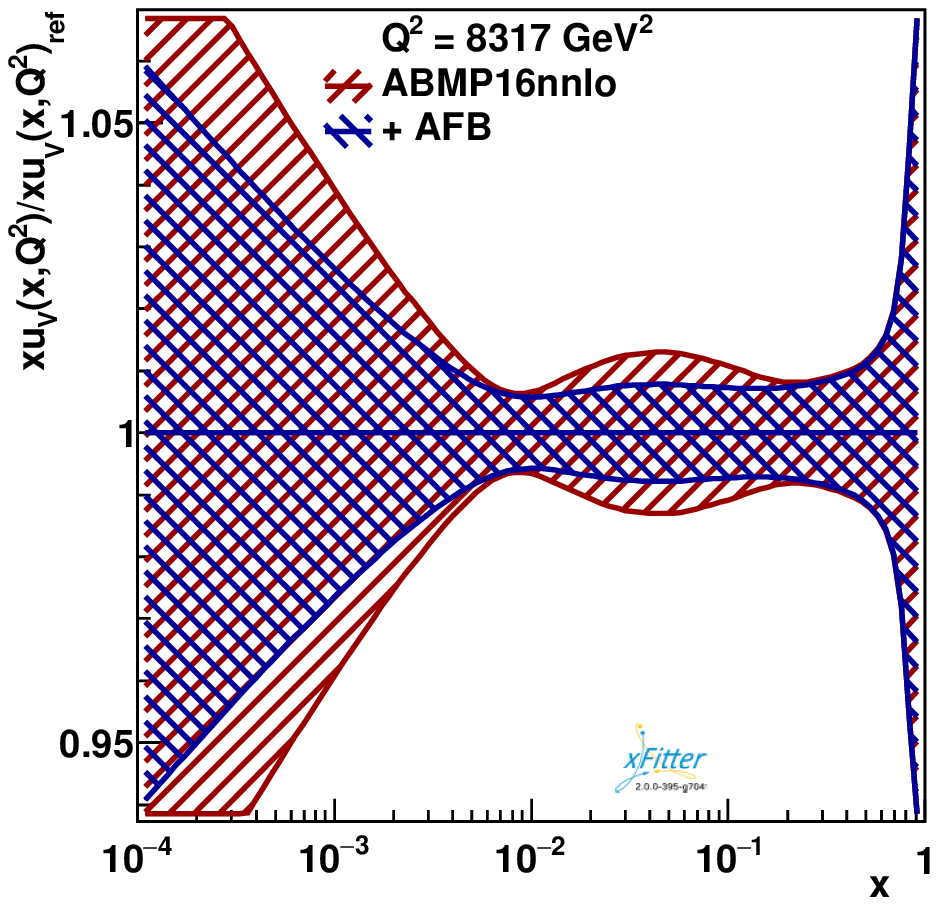}\includegraphics[width=0.25\textwidth]{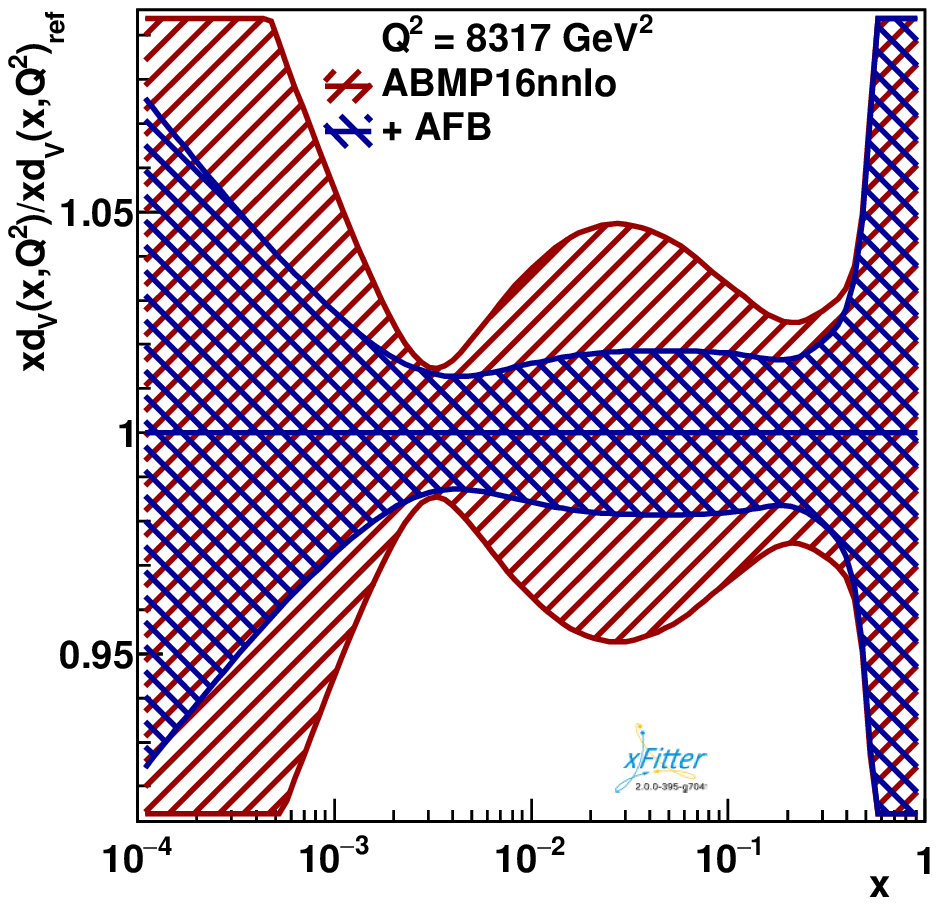}\includegraphics[width=0.25\textwidth]{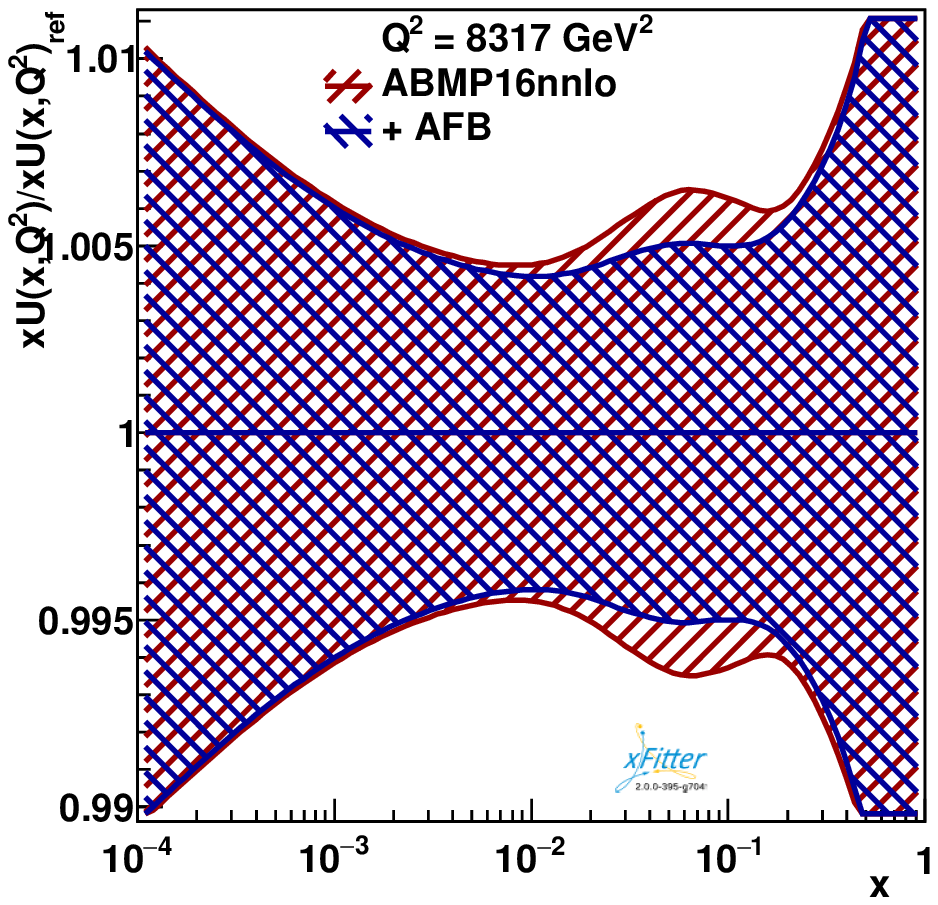}\includegraphics[width=0.25\textwidth]{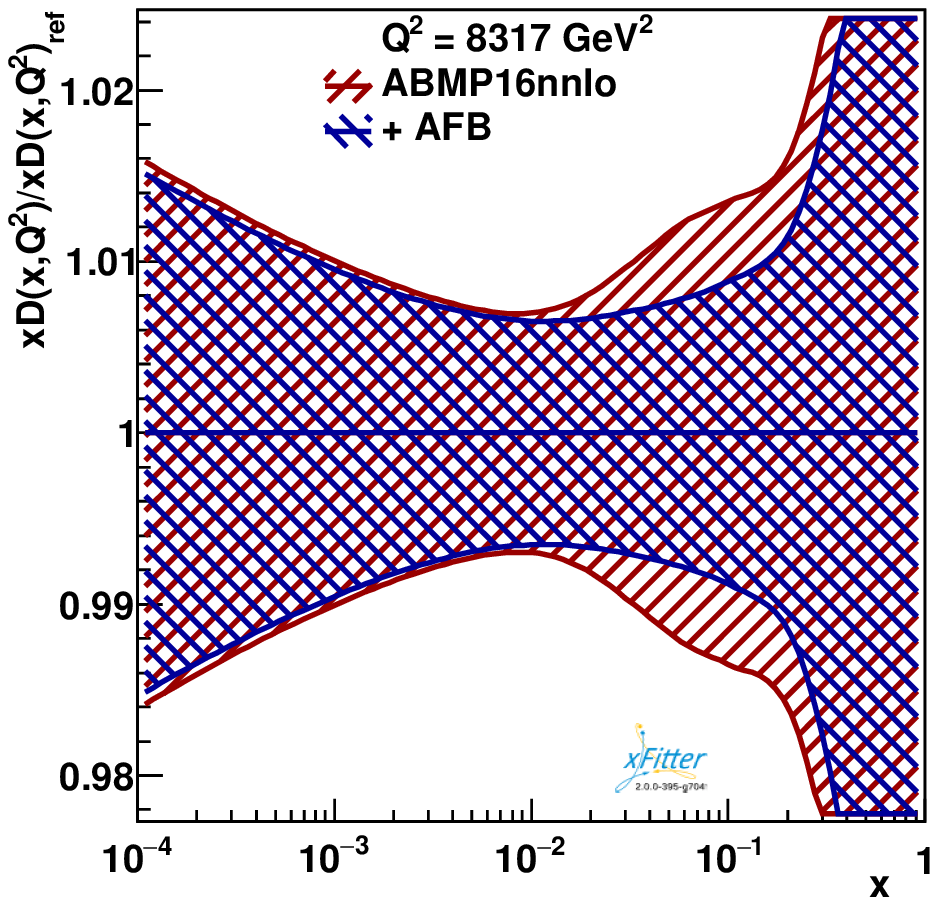}\\\includegraphics[width=0.25\textwidth]{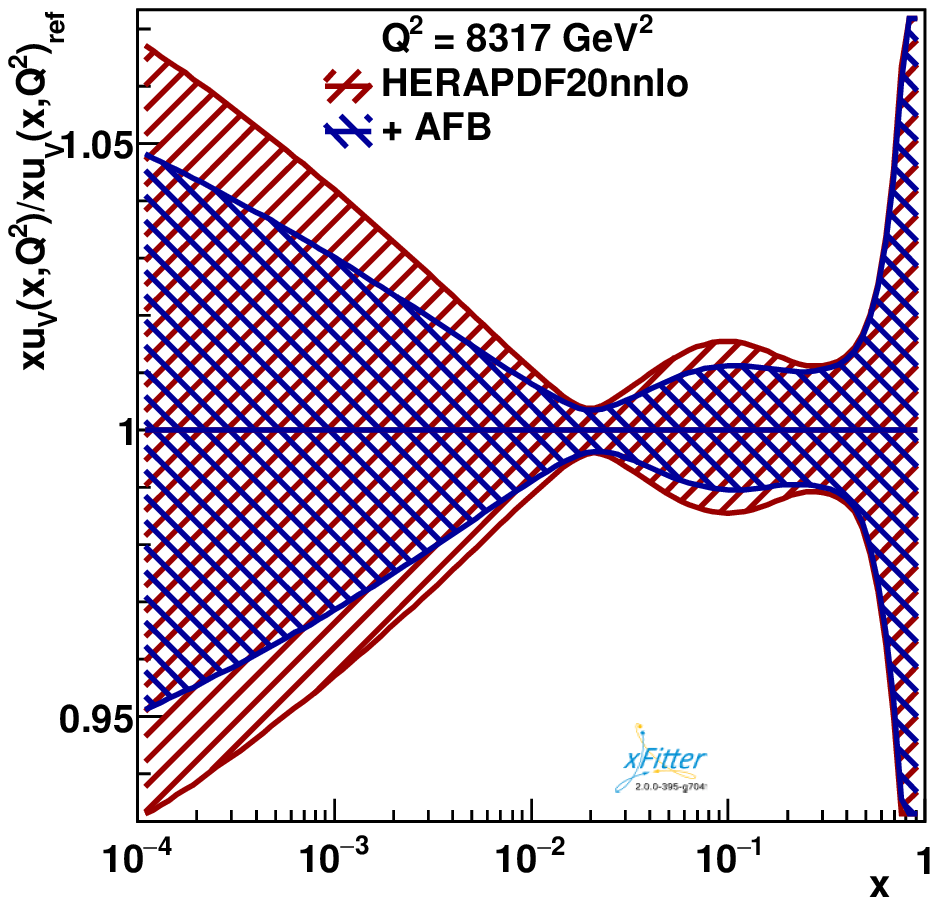}\includegraphics[width=0.25\textwidth]{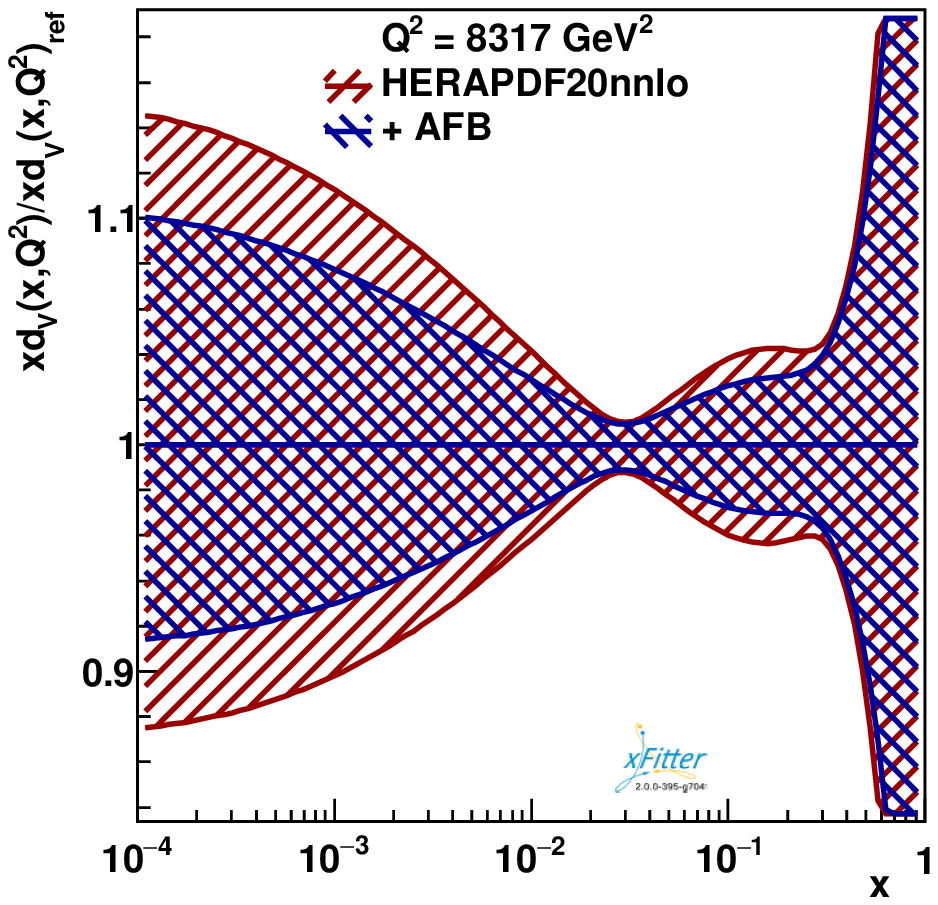}\includegraphics[width=0.25\textwidth]{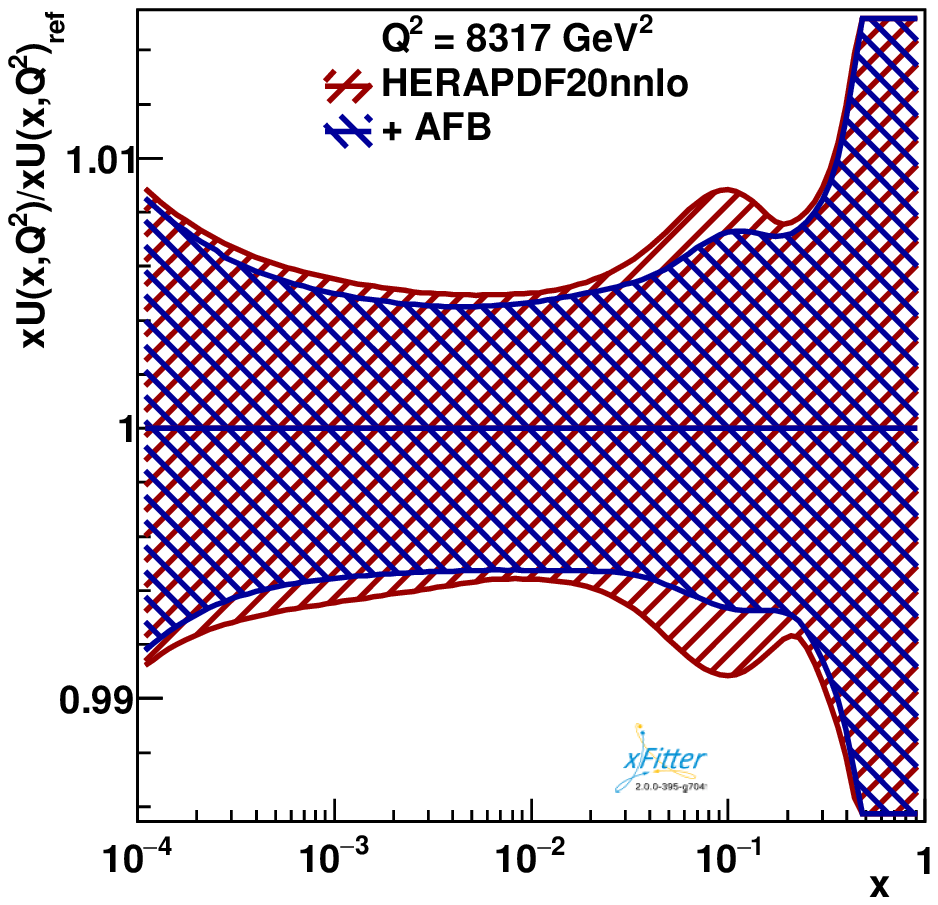}\includegraphics[width=0.25\textwidth]{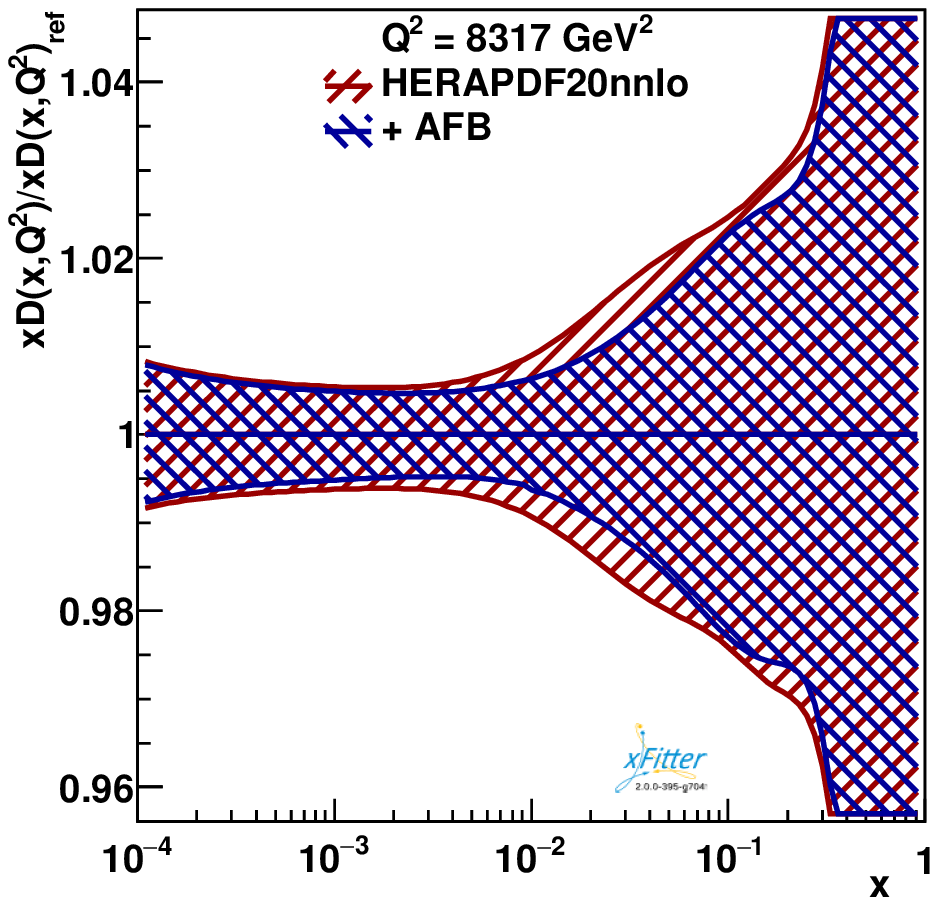}
\caption{Original and profiled distributions for the normalised ratios of (left to right) $u$-valence, $d$-valence, $u$-sea and $d$-sea quarks, obtained with $A_{\rm{FB}}^*$ pseudodata at L = 300 fb$^{-1}$. Distributions are shown for the PDF sets (top to bottom) NNPDF3.1nnlo, MMHT2014nnlo, ABMP16nnlo and HERA2.0nnlo.}
\label{fig:prof_PDFs}
\end{center}
\end{figure}

In a similar way, in Fig.~\ref{fig:prof_PDFs} we show the profiled curves for the PDF sets (rows from top to bottom) NNPDF3.1nnlo (Hessian set)~\cite{Ball:2017nwa}, MMHT2014nnlo~\cite{Harland-Lang:2014zoa}, ABMP16nnlo~\cite{Alekhin:2017kpj} and HERA2.0nnlo (EIG)~\cite{Abramowicz:2015mha}, obtained including $A_{\rm{FB}}$ data with a projected statistical error corresponding to an integrated luminosity of 300 fb$^{-1}$.
The ABMP16nnlo and HERA2.0nnlo sets appear to be the most sensitive to the new data, especially in the distributions of $u$-valence and $d$-valence quarks in the region of low and intermediate $x$, while less improvement is visible in the sea quarks distributions.
The NNPDF3.1nnlo and MMHT2014nnlo sets also show some improvement in the in the region of low to intermediate $x$ for the valence quarks distribution, while also a visible improvement is observed in the $u$-sea (especially for NNPDF3.1nnlo) and in the $d$-sea (especially for MMHT2014nnlo) error bands.

\begin{figure}[h]
\begin{center}
\includegraphics[width=0.33\textwidth]{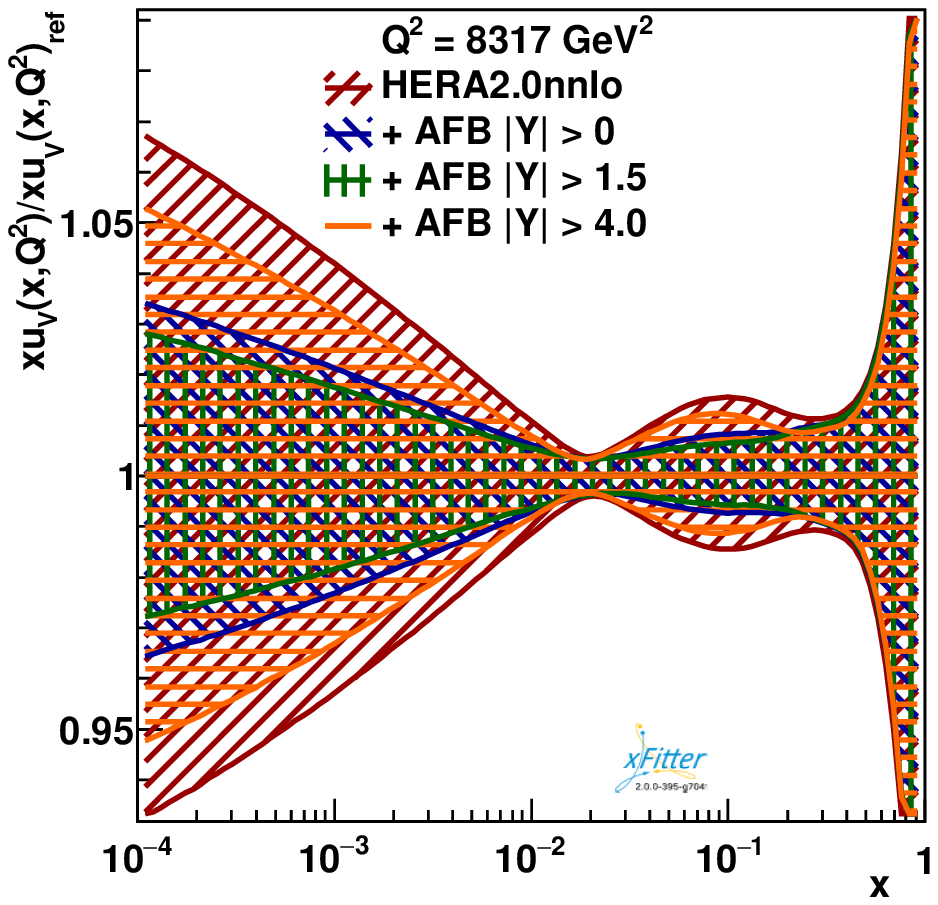}\includegraphics[width=0.33\textwidth]{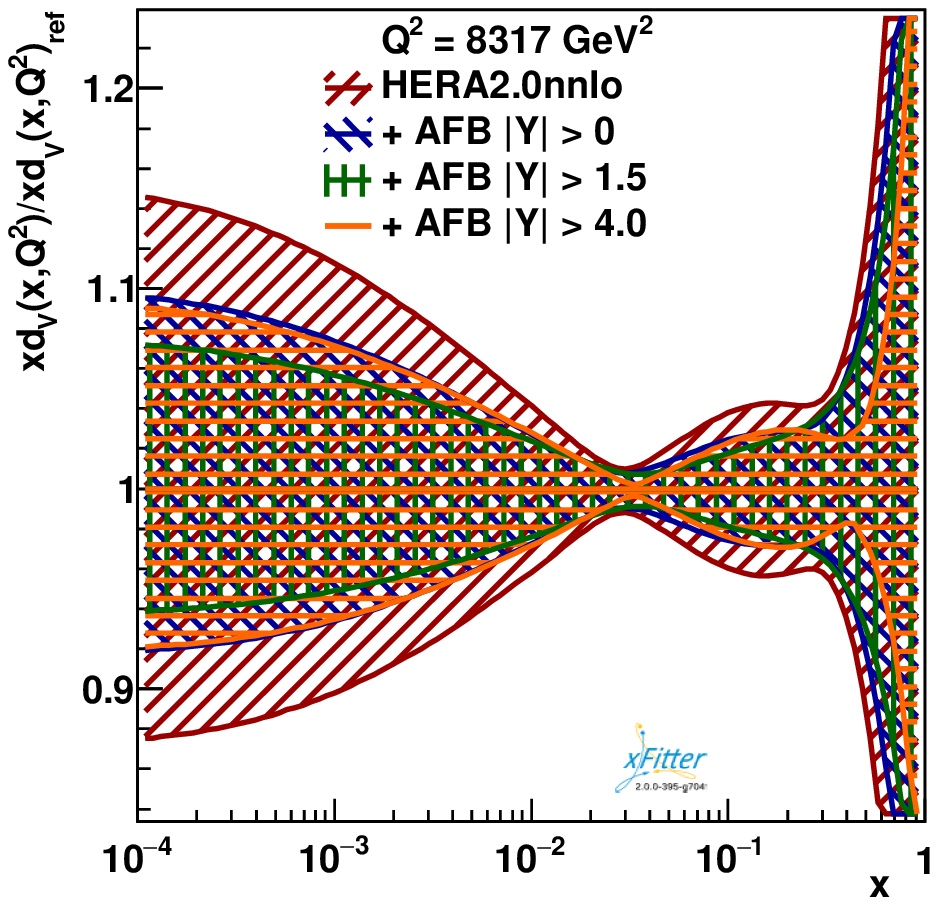}\includegraphics[width=0.33\textwidth]{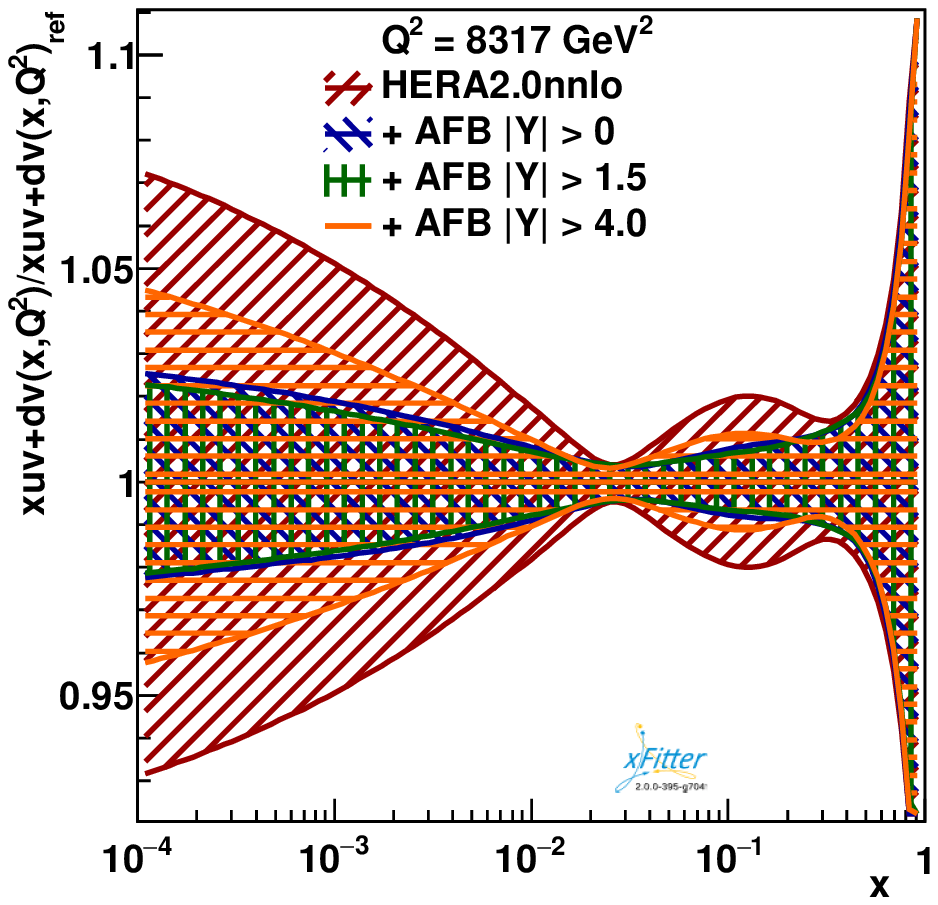}
\caption{Original (red) and profiled distributions for the normalised ratios of $u$-valence, $d$-valence and $(u+d)$-valence, obtained with $A_{\rm{FB}}^*$ pseudodata at L = 3000 fb$^{-1}$ applying rapidity cuts of $|y_{\ell\ell}| > 0$ (blue), $|y_{\ell\ell}| > 1.5$ (green) and $|y_{\ell\ell}| > 4$ (orange) respectively.
In the latter case the analysis is performed at LO and acceptance region of the detector has been enlarged up to $|\eta_\ell| < 5$.}
\label{fig:prof_HERA_Y_cut}
\end{center}
\end{figure}

In Fig.~\ref{fig:prof_HERA_Y_cut} we show the effects of the application of low rapidity cuts on the data on the profiled error bands of the HERA2.0nnlo PDF set, using $A_{\rm{FB}}$ data with a statistical error corresponding to an integrated luminosity of 3000 fb$^{-1}$.
We display the results for the distribution of $u$-valence, $d$-valence and their combination $(u+d)$-valence quarks, using data where rapidity cuts of $|y_{\ell\ell}| > 0$ (blue), $|y_{\ell\ell}| > 1.5$ (green) and $|y_{\ell\ell}| > 4.0$ (orange) have been applied.
In the latter case the analysis is performed using $A_{\rm{FB}}^*$ predictions at LO and pseudodata generated in the invariant mass range $80~{\rm GeV} < M_{\ell\ell} < 200~{\rm GeV}$ and with a bin size of 1 GeV.
Furthermore, in order to explore the very high rapidity region, we have extended the detector acceptance region up to pseudorapidity $|\eta_\ell| < 5$.
When comparing the profiled error bands obtained with $|y_{\ell\ell}| > 0$ and $|y_{\ell\ell}| > 1.5$, we observe a remarkable improvement in the distribution of the valence quarks, especially in the region of low $x$, which is then reflected in the combination $(u+d)$-valence.
On the other hand, the curve obtained with a rapidity cut $|y_{\ell\ell}| > 4.0$ shows an overall weaker profiling compared to the previous cases.
This is mostly due to the reduced statistic consequence of the stronger phase space cut.
However here the data allows to access higher $x$ and we can appreciate a reduction of the uncertainty error bands concentrated in that region.
We therefore find that new PDF sensitivity arises from the di-lepton mass and rapidity spectra of the AFB (which has traditionally been used for determination of the weak mixing angle, see e.g.~\cite{Aad:2015uau,Chatrchyan:2011ya,Aaij:2015lka,Bodek:2016olg,ATLAS:2018gqq,Bodek:2018sin,Sirunyan:2018swq} and references therein).

\section{Conclusions}

The new data from the LHC Runs 2 and the upcoming Run 3 and HL-LHC stages will provide important information to constrain non-perturbative QCD effects from PDFs.
DY data in particular can be exploited both in the form of cross section and asymmetry.
The latter contains a complementary information with respect to the cross section and it also encodes information on the lepton polar angle, or pseudorapidity.

We have considered the constraints on PDFs from the inclusion of $A_{\rm{FB}}$ data, implementing this observable in the {\tt{xFitter}} framework and obtaining profiled curves to assess the impact of the new data.
Suitable pseudodata have been generated for such purpose for various PDF sets, projecting current and future statistical precision of the measurements with integrated luminosities of 30 fb$^{-1}$, 300 fb$^{-1}$ and 3000 fb$^{-1}$.

We have observed a significant improvement of PDF uncertainties, especially in the valence quarks distribution in the low and intermediate $x$ region.
Some amelioration has been also obtained in the sea quarks distributions when employing high statistics data.
We have evidenced substantial differences in the results obtained for the different PDF sets.

Furthermore, we have studied the impact of rapidity cuts on the di-lepton system, and we noted that by these means we can further improve PDF distributions in the high $x$ region.

\bibliographystyle{JHEP}

\bibliography{bib}

\providecommand{\href}[2]{#2}\begingroup\raggedright\begin{thebibliography}{10}

\bibitem{Accomando:2018nig}
E.~Accomando, J.~Fiaschi, F.~Hautmann and S.~Moretti, \emph{{Neutral current
  forward backward asymmetry: from $\theta _W$ to PDF determinations}},
  \href{https://doi.org/10.1140/epjc/s10052-018-6120-6}{\emph{Eur. Phys. J.}
  {\bfseries C78} (2018) 663}
  [\href{https://arxiv.org/abs/1805.09239}{{\ttfamily 1805.09239}}].

\bibitem{Accomando:2017scx}
E.~Accomando, J.~Fiaschi, F.~Hautmann and S.~Moretti, \emph{{Constraining
  Parton Distribution Functions from Neutral Current Drell-Yan Measurements}},
  \href{https://doi.org/10.1103/PhysRevD.98.013003}{\emph{Phys. Rev.}
  {\bfseries D98} (2018) 013003}
  [\href{https://arxiv.org/abs/1712.06318}{{\ttfamily 1712.06318}}].

\bibitem{Alekhin:2014irh}
S.~Alekhin et~al., \emph{{HERAFitter}},
  \href{https://doi.org/10.1140/epjc/s10052-015-3480-z}{\emph{Eur. Phys. J.}
  {\bfseries C75} (2015) 304}
  [\href{https://arxiv.org/abs/1410.4412}{{\ttfamily 1410.4412}}].

\bibitem{Abdolmaleki:2019qmq}
E.~Accomando, J.~Fiaschi, F.~Hautmann, S.~Moretti, H.~the xFitter
  Developers'~team: Abdolmaleki, V.~Bertone et~al., \emph{{PDF Profiling Using
  the Forward-Backward Asymmetry in Neutral Current Drell-Yan Production}}
  [\href{https://arxiv.org/abs/1907.07727}{{\ttfamily 1907.07727}}].

\bibitem{Dittmar:1996my}
M.~Dittmar, \emph{{Neutral current interference in the TeV region: The
  Experimental sensitivity at the LHC}},
  \href{https://doi.org/10.1103/PhysRevD.55.161}{\emph{Phys. Rev.} {\bfseries
  D55} (1997) 161} [\href{https://arxiv.org/abs/hep-ex/9606002}{{\ttfamily
  hep-ex/9606002}}].

\bibitem{Rizzo:2009pu}
T.~G. Rizzo, \emph{{Indirect Searches for Z$^\prime$-like Resonances at the
  LHC}}, \href{https://doi.org/10.1088/1126-6708/2009/08/082}{\emph{JHEP}
  {\bfseries 08} (2009) 082} [\href{https://arxiv.org/abs/0904.2534}{{\ttfamily
  0904.2534}}].

\bibitem{Accomando:2015cfa}
E.~Accomando, A.~Belyaev, J.~Fiaschi, K.~Mimasu, S.~Moretti and
  C.~Shepherd-Themistocleous, \emph{{Forward-backward asymmetry as a discovery
  tool for Z$^\prime$ bosons at the LHC}},
  \href{https://doi.org/10.1007/JHEP01(2016)127}{\emph{JHEP} {\bfseries 01}
  (2016) 127} [\href{https://arxiv.org/abs/1503.02672}{{\ttfamily
  1503.02672}}].

\bibitem{Accomando:2016tah}
E.~Accomando, J.~Fiaschi, F.~Hautmann, S.~Moretti and C.~H.
  Shepherd-Themistocleous, \emph{{Photon-initiated production of a dilepton
  final state at the LHC: Cross section versus forward-backward asymmetry
  studies}}, \href{https://doi.org/10.1103/PhysRevD.95.035014}{\emph{Phys.
  Rev.} {\bfseries D95} (2017) 035014}
  [\href{https://arxiv.org/abs/1606.06646}{{\ttfamily 1606.06646}}].

\bibitem{Accomando:2016ehi}
E.~Accomando, J.~Fiaschi, F.~Hautmann, S.~Moretti and C.~H.
  Shepherd-Themistocleous, \emph{{The effect of real and virtual photons in the
  di-lepton channel at the LHC}},
  \href{https://doi.org/10.1016/j.physletb.2017.04.025}{\emph{Phys. Lett.}
  {\bfseries B770} (2017) 1}
  [\href{https://arxiv.org/abs/1612.08168}{{\ttfamily 1612.08168}}].

\bibitem{Alwall:2014hca}
J.~Alwall, R.~Frederix, S.~Frixione, V.~Hirschi, F.~Maltoni, O.~Mattelaer
  et~al., \emph{{The automated computation of tree-level and next-to-leading
  order differential cross sections, and their matching to parton shower
  simulations}}, \href{https://doi.org/10.1007/JHEP07(2014)079}{\emph{JHEP}
  {\bfseries 07} (2014) 079} [\href{https://arxiv.org/abs/1405.0301}{{\ttfamily
  1405.0301}}].

\bibitem{Carli:2010rw}
T.~Carli, D.~Clements, A.~Cooper-Sarkar, C.~Gwenlan, G.~P. Salam, F.~Siegert
  et~al., \emph{{A posteriori inclusion of parton density functions in NLO QCD
  final-state calculations at hadron colliders: The APPLGRID Project}},
  \href{https://doi.org/10.1140/epjc/s10052-010-1255-0}{\emph{Eur. Phys. J.}
  {\bfseries C66} (2010) 503}
  [\href{https://arxiv.org/abs/0911.2985}{{\ttfamily 0911.2985}}].

\bibitem{Bertone:2014zva}
V.~Bertone, R.~Frederix, S.~Frixione, J.~Rojo and M.~Sutton, \emph{{aMCfast:
  automation of fast NLO computations for PDF fits}},
  \href{https://doi.org/10.1007/JHEP08(2014)166}{\emph{JHEP} {\bfseries 08}
  (2014) 166} [\href{https://arxiv.org/abs/1406.7693}{{\ttfamily 1406.7693}}].

\bibitem{Collins:1977iv}
J.~C. Collins and D.~E. Soper, \emph{{Angular Distribution of Dileptons in
  High-Energy Hadron Collisions}},
  \href{https://doi.org/10.1103/PhysRevD.16.2219}{\emph{Phys. Rev.} {\bfseries
  D16} (1977) 2219}.

\bibitem{Khachatryan:2014fba}
{\scshape CMS} collaboration, \emph{{Search for physics beyond the standard
  model in dilepton mass spectra in proton-proton collisions at $ \sqrt{s}=8 $
  TeV}}, \href{https://doi.org/10.1007/JHEP04(2015)025}{\emph{JHEP} {\bfseries
  04} (2015) 025} [\href{https://arxiv.org/abs/1412.6302}{{\ttfamily
  1412.6302}}].

\bibitem{Hamberg:1990np}
R.~Hamberg, W.~L. van Neerven and T.~Matsuura, \emph{{A complete calculation of
  the order $\alpha_s^{2}$ correction to the Drell-Yan $K$ factor}},
  \href{https://doi.org/10.1016/S0550-3213(02)00814-3,
  10.1016/0550-3213(91)90064-5}{\emph{Nucl. Phys.} {\bfseries B359} (1991)
  343}.

\bibitem{Harlander:2002wh}
R.~V. Harlander and W.~B. Kilgore, \emph{{Next-to-next-to-leading order Higgs
  production at hadron colliders}},
  \href{https://doi.org/10.1103/PhysRevLett.88.201801}{\emph{Phys. Rev. Lett.}
  {\bfseries 88} (2002) 201801}
  [\href{https://arxiv.org/abs/hep-ph/0201206}{{\ttfamily hep-ph/0201206}}].

\bibitem{Paukkunen:2014zia}
H.~Paukkunen and P.~Zurita, \emph{{PDF reweighting in the Hessian matrix
  approach}}, \href{https://doi.org/10.1007/JHEP12(2014)100}{\emph{JHEP}
  {\bfseries 12} (2014) 100} [\href{https://arxiv.org/abs/1402.6623}{{\ttfamily
  1402.6623}}].

\bibitem{Dulat:2015mca}
S.~Dulat, T.-J. Hou, J.~Gao, M.~Guzzi, J.~Huston, P.~Nadolsky et~al.,
  \emph{{New parton distribution functions from a global analysis of quantum
  chromodynamics}},
  \href{https://doi.org/10.1103/PhysRevD.93.033006}{\emph{Phys. Rev.}
  {\bfseries D93} (2016) 033006}
  [\href{https://arxiv.org/abs/1506.07443}{{\ttfamily 1506.07443}}].

\bibitem{Ball:2017nwa}
{\scshape NNPDF} collaboration, \emph{{Parton distributions from high-precision
  collider data}},
  \href{https://doi.org/10.1140/epjc/s10052-017-5199-5}{\emph{Eur. Phys. J.}
  {\bfseries C77} (2017) 663}
  [\href{https://arxiv.org/abs/1706.00428}{{\ttfamily 1706.00428}}].

\bibitem{Harland-Lang:2014zoa}
L.~A. Harland-Lang, A.~D. Martin, P.~Motylinski and R.~S. Thorne, \emph{{Parton
  distributions in the LHC era: MMHT 2014 PDFs}},
  \href{https://doi.org/10.1140/epjc/s10052-015-3397-6}{\emph{Eur. Phys. J.}
  {\bfseries C75} (2015) 204}
  [\href{https://arxiv.org/abs/1412.3989}{{\ttfamily 1412.3989}}].

\bibitem{Alekhin:2017kpj}
S.~Alekhin, J.~Blümlein, S.~Moch and R.~Placakyte, \emph{{Parton distribution
  functions, $\alpha_s$, and heavy-quark masses for LHC Run II}},
  \href{https://doi.org/10.1103/PhysRevD.96.014011}{\emph{Phys. Rev.}
  {\bfseries D96} (2017) 014011}
  [\href{https://arxiv.org/abs/1701.05838}{{\ttfamily 1701.05838}}].

\bibitem{Abramowicz:2015mha}
{\scshape ZEUS, H1} collaboration, \emph{{Combination of measurements of
  inclusive deep inelastic ${e^{\pm }p}$ scattering cross sections and QCD
  analysis of HERA data}},
  \href{https://doi.org/10.1140/epjc/s10052-015-3710-4}{\emph{Eur. Phys. J.}
  {\bfseries C75} (2015) 580}
  [\href{https://arxiv.org/abs/1506.06042}{{\ttfamily 1506.06042}}].

\bibitem{Aad:2015uau}
{\scshape ATLAS} collaboration, \emph{{Measurement of the forward-backward
  asymmetry of electron and muon pair-production in $pp$ collisions at
  $\sqrt{s}$ = 7 TeV with the ATLAS detector}},
  \href{https://doi.org/10.1007/JHEP09(2015)049}{\emph{JHEP} {\bfseries 09}
  (2015) 049} [\href{https://arxiv.org/abs/1503.03709}{{\ttfamily
  1503.03709}}].

\bibitem{Chatrchyan:2011ya}
{\scshape CMS} collaboration, \emph{{Measurement of the weak mixing angle with
  the Drell-Yan process in proton-proton collisions at the LHC}},
  \href{https://doi.org/10.1103/PhysRevD.84.112002}{\emph{Phys. Rev.}
  {\bfseries D84} (2011) 112002}
  [\href{https://arxiv.org/abs/1110.2682}{{\ttfamily 1110.2682}}].

\bibitem{Aaij:2015lka}
{\scshape LHCb} collaboration, \emph{{Measurement of the forward-backward
  asymmetry in $Z/\gamma^{\ast} \rightarrow \mu^{+}\mu^{-}$ decays and
  determination of the effective weak mixing angle}},
  \href{https://doi.org/10.1007/JHEP11(2015)190}{\emph{JHEP} {\bfseries 11}
  (2015) 190} [\href{https://arxiv.org/abs/1509.07645}{{\ttfamily
  1509.07645}}].

\bibitem{Bodek:2016olg}
A.~Bodek, J.~Han, A.~Khukhunaishvili and W.~Sakumoto, \emph{{Using Drell–Yan
  forward–backward asymmetry to reduce PDF uncertainties in the measurement
  of electroweak parameters}},
  \href{https://doi.org/10.1140/epjc/s10052-016-3958-3}{\emph{Eur. Phys. J.}
  {\bfseries C76} (2016) 115}
  [\href{https://arxiv.org/abs/1507.02470}{{\ttfamily 1507.02470}}].

\bibitem{ATLAS:2018gqq}
{\scshape ATLAS} collaboration, \emph{{Measurement of the effective leptonic
  weak mixing angle using electron and muon pairs from $Z$-boson decay in the
  ATLAS experiment at $\sqrt s = 8$ TeV}}, {\emph{ATLAS-CONF-2018-037} (2018)
  }.

\bibitem{Bodek:2018sin}
{\scshape CMS} collaboration, \emph{{Measurement of the effective weak mixing
  angle $sin^2\theta^{lept}_{eff}$ from the forward-backward asymmetry of
  Drell-Yan events at CMS}},  in \emph{{13th Conference on the Intersections of
  Particle and Nuclear Physics (CIPANP 2018) Palm Springs, California, USA, May
  29-June 3, 2018}}, 2018, [\href{https://arxiv.org/abs/1808.03170}{{\ttfamily
  1808.03170}}].

\bibitem{Sirunyan:2018swq}
{\scshape CMS} collaboration, \emph{{Measurement of the weak mixing angle using
  the forward-backward asymmetry of Drell-Yan events in pp collisions at 8
  TeV}}, \href{https://doi.org/10.1140/epjc/s10052-018-6148-7}{\emph{Eur. Phys.
  J.} {\bfseries C78} (2018) 701}
  [\href{https://arxiv.org/abs/1806.00863}{{\ttfamily 1806.00863}}].

\end{thebibliography}\endgroup

\end{document}